\documentstyle[pre,aps,floats,epsf,amssymb,psfig,epsfig]{revtex}
\begin{document}
\draft

\twocolumn[\hsize\textwidth\columnwidth\hsize\csname@twocolumnfalse%
\endcsname

\title {Vehicular Traffic: A System of Interacting Particles Driven Far From Equilibrium$^{\ast}$}

\author{Debashish Chowdhury}
\address{Department of Physics, Indian Institute of Technology, Kanpur 208016, India$^{\dagger}$}

\address{Institute for Theoretical Physics, University of Cologne, D-50923 K\"oln, Germany}

\author{Ludger Santen and Andreas Schadschneider}

\address{Institute for Theoretical Physics, University of Cologne, D-50923 K\"oln, Germany}

\maketitle


\begin{abstract}
In recent years statistical physicists have developed {\it discrete} 
"particle-hopping" models of vehicular traffic, usually formulated 
in terms of {\it cellular automata}, which are similar to the 
microscopic models of interacting charged particles in the presence 
of an external electric field. Concepts and techniques of 
non-equilibrium statistical mechanics are being used to understand 
the nature of the steady states and fluctuations in these so-called 
"microscopic" models. In this brief review we explain, primarily to 
the nonexperts, these models and the physical implications of the results.
\end{abstract}

]

\newpage 

\section{Introduction}

Are you surprised to see an article on vehicular traffic  
in this special section of current science where physicists are 
supposed to report on some recent developments in the area of 
dynamics of nonequilibrium statistical systems? Aren't civil 
engineers (or, more specifically, traffic engineers) expected to 
work on traffic? Solving traffic problems would become easier 
if one knows the fundamental laws governing traffic flow and 
traffic jam. For almost half a century physicists have been 
trying to develope a theoretical framework of traffic science 
extending concepts and techniques of statistical physics
\cite{hg,gazis,ph,proc1,proc2,helbsci}. The main aim of this brief 
review is to show how these attempts, particularly the recent 
ones, have led to deep insight in this frontier area of 
inter-disciplinary research. 

The dynamical phases of systems driven far from equilibrium are 
counterparts of the stable phases of systems in equilibrium. 
Let us first pose some of the questions that statistical physicists 
have been addressing in order to discover the {\it fundamental laws} 
governing vehicular traffic. For example,\\ 
(i) What are the various {\it dynamical phases} of traffic? Does 
traffic exhibit phase-coexistence, phase transition, criticality or 
self-organized criticality and, if so, under which circumstances?\\
(ii) What is the nature of {\it fluctuations around the steady-states} 
of traffic?\\ 
(iii) If the initial state is far from a stationary state of the 
driven system, how does it {\it evolve with time} to reach a 
truely steady-state? \\ 
(iv) What are the effects of {\it quenched disorder} (i.e., 
time-independent disorder) on the answers to the questions posed 
in (i)-(iii) above?\\
The microscopic models of vehicular traffic can find {\it 
practical applications} in on-line traffic control systems 
as well as in the planning and design of transportation network. 

There are two different conceptual frameworks for modellig vehicular 
traffic. In the "coarse-grained" fluid-dynamical description, the 
traffic is viewed as a compressible fluid formed by the vehicles but 
these individual vehicles do not appear explicitly in the theory. In 
contrast, in the "microscopic" models traffic is treated as {\it a  
system of interacting "particles" driven far from equilibrium} 
where attention is explicitly focussed on individual vehicles 
each of which is represented by a "particle"; the nature of the 
interactions among these particles is determined by the way the 
vehicles influence each others' movement. Unlike the particles 
in a gas, a driver is an intelligent agent who can "think", make 
individual decisions and "learn" from experience. Nevertheless, 
many general phenomena in traffic can be explained in general terms 
with these models provided the behavioural effects of the drivers 
are captured by only a few phenomenological parameters. 

The conceptual basis of the older theoretical approaches are 
explained briefly in section II. Most of the "microscopic" models 
developed in the recent years are "particle-hopping" models which 
are usually formulated using the language of cellular automata 
(CA)~\cite{wolfram}. The Nagel-Schreckenberg (NaSch)~\cite{ns} 
model and the Biham-Middleton-Levine (BML) ~\cite{bml} model, which 
are the most popular CA models of traffic on idealized highways 
and cities, respectively, have been extended by several authors to 
develope more realistic models. Some of the most interesting aspects 
of these recent developments are discussed in the long sections III 
and IV. The similarities between various particle-hopping models of 
traffic and some other models of systems, which are also far from 
equilibrium, are pointed out in section V followed by the concluding 
section VI. 

\section{Older theories of vehicular traffic}

\subsection{Fluid-dynamical Theories of vehicular traffic} 

In traffic engineering, the {\it fundamental diagram} depicts the 
relation between density $c$ and the {\bf flux} $J$, which is 
defined as the number of vehicles crossing a detector site per 
unit time \cite{may90}. Because of the conservation of vehicles, 
the local density $c(x;t)$ and local flux $J(x;t)$ satisfy the 
equation of continuity which is the analogue of the equation of 
continuity in the hydrodynamic theories of fluids. In the early 
works\cite{lw} it was assumed (i) that the flux (or, equivalently, 
the velocity) is a function of  the density and (ii) that, 
following any change in the local density, the local speed 
instantaneously relaxes to a magnitude consistent with the new 
density at the same location. However, for a more realistic 
description of traffic, in the recent fluid-dynamical treatments 
\cite{kernetal,hongetal,leeetal} of traffic an additional equation 
(the analogue of the Navier-Stokes equation for fluids), which 
describes the time-dependence of the velocity $V(x;t)$, has been 
considered. This approach, however, has its limitations; for 
example, viscosity of traffic is not a directly measureable quantity.

\subsection{Kinetic theory of vehicular traffic} 

In the kinetic theory of traffic, one begins with the basic quantity 
$g(x,v,w;t) ~dx ~dv ~dw$ which is the number of vehicles, at time 
$t$, located between $x$ and $x+dx$, having {\it actual} velocity 
between $v$ and $v + dv$ and {\it desired} velocity between $w$ 
and $w + dw$. In this approach, the fundamental dynamical equation 
is the analogue of the Boltzmann equation in the kinetic theory of 
gases\cite{ph}. Assuming reasonable forms of "relaxation" and 
"interaction", the problem of traffic is reduced to that of solving 
the Boltzmann-like equation, a formidable task, indeed 
\cite{helbkin,wagkin,nagakin}!

\subsection{Car-following theories of vehicular traffic} 

In the car-following theories one writes, for each individual vehicle,
an equation of motion which is the analogue of the Newton's equation
for each individual particle in a system of interacting classical
particles. In Newtonian mechanics, the acceleration may be regarded as
the {\it response} of the particle to the {\it stimulus} it receives
in the form of force which includes both the external force as well
as those arising from its interaction with all the other particles in
the system. Therefore, the basic philisophy of the car-following
theories\cite{hg,gazis} can be summarized by the equation
\begin{equation}
[Response]_n \propto [Stimulus]_n
\label{car-follow}
\end{equation}
for the $n$-th vehicle ($n = 1,2,...$). The constant of proportionality 
in the equation (\ref{car-follow}) can be interpreted as a measure of 
the sensitivity coefficient of the driver; it indicates how strongly 
the driver responds to unit stimulus. Each driver can respond to the 
surrounding traffic conditions only by accelerating or decelerating 
the vehicle. The stimulus and the sensitivity factor are assumed to be 
functions of the position and speed of the vehicle under consideration 
and those of its leading vehicle. Different forms of the equations of 
motion of the vehicles in the different versions of the car-following 
models arise from the differences in their postulates regarding the 
nature of the stimulus.  In general, the dynamical equations for the 
vehicles in the car-following theories are coupled non-linear 
differential equations
~\cite{bandoetal,komatsu,hayanaka,mason,sugiyama,nagatani,wagner} 
and thus, in this "microscopic" approach, the problem of traffic 
flow reduces to problems of nonlinear dynamics. 

\section{Cellular-automata Models of Highway-traffic}

In the car-following models space is treated as a continuum and time 
is represented by a continuous variable $t$ while velocities and 
accelerations of the vehicles are also real variables. However, most 
often, for numerical manipulations of the differential equations of 
the car-following models, one needs to discretize the continuous 
variables with appropriately chosen grids. In contrast, in the CA 
models of traffic not only time but also the position, speed, and 
acceleration of the vehicles are treated as {\it discrete} variables. 
In this approach, a lane is represented by a one-dimensional lattice. 
Each of the lattice sites represents a "cell" which can be either 
empty or occupied by at most one "vehicle" at a given instant of time 
(see fig.1). 
\begin{figure}[h]
 \centerline{\psfig{figure=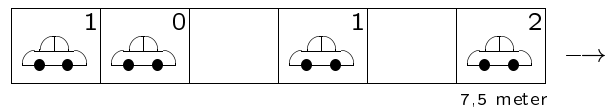,bbllx=50pt,bblly=420pt,bburx=550pt,bbury=540pt,height=2cm}}
\caption{\protect{ A typical configuration in a CA model. The 
number in the upper right corner is the speed of the vehicle. }}
\label{fig1}
\end{figure}

At each {\it discrete time} step $t \rightarrow t+1$, the state of the 
system is updated following a well defined prescription.

\subsection{The Nagel-Schreckenberg model of highway traffic:}

In the NaSch model, the speed $V$ of each vehicle can take one of 
the $V_{max}+1$ allowed {\it integer} values $V=0,1,...,V_{max}$. 
Suppose, $X_n$ and $V_n$ denote the position and speed, respectively, 
of the $n$-th vehicle. Then, $d_n = X_{n+1}-X_n$, is the gap in 
between the $n$-th vehicle and the vehicle in front of it at time $t$. 
At each time step $t \rightarrow t+1$, the arrangement of the $N$ 
vehicles on a finite lattice of length $L$ is updated {\it in 
parallel} according to the following "rules":

\noindent {\it Step 1: Acceleration.} If, $ V_n < V_{max}$, the 
speed of the $n$-th vehicle is increased by one, but $V_n$ remains 
unaltered if $V_n = V_{max}$, i.e., $V_n \rightarrow min(V_n+1,V_{max})$. 

\noindent{\it Step 2: Deceleration (due to other vehicles).} If 
$d_n \le V_n$, the speed of the $n$-th vehicle is reduced to $d_n-1$, 
i.e., $V_n \rightarrow min(V_n,d_n-1)$.  

\noindent{\it Step 3: Randomization.} If $V_n > 0$, the speed of the $n$-th
vehicle is decreased randomly by unity with probability $p$ but $V_n$ 
does not change if $V_n = 0$, i.e., 
$V_n \rightarrow max(V_n-1,0)$ with probability $p$.  

\noindent{\it Step 4: Vehicle movement.} Each vehicle is moved forward so
that $X_n \rightarrow  X_n + V_n$.

The NaSch model is a minimal model in the sense that all the four 
steps are necessary to reproduce the basic features of real traffic; 
however, additional rules need to be formulated to capture more 
complex situations. 
The step 1 reflects the general tendency of the drivers to drive 
as fast as possible, if allowed to do so, without crossing the 
maximum speed limit. The step 2 is intended to avoid collision 
between the vehicles. The randomization in step 3 takes into 
account the different behavioural patterns of the individual drivers, 
especially, nondeterministic acceleration as well as overreaction 
while slowing down; this is crucially important for the spontaneous 
formation of traffic jams. So long as $p \neq 0$, the NaSch model 
may be regarded as stochastic CA~\cite{wolfram}. For a realistic 
description of highway traffic \cite{ns}, the typical length of 
each cell should be about $7.5$m and each time step should correspond 
to approximately $1$ sec of real time when $V_{max} = 5$. 

\begin{figure}[h]
 \centerline{\psfig{figure=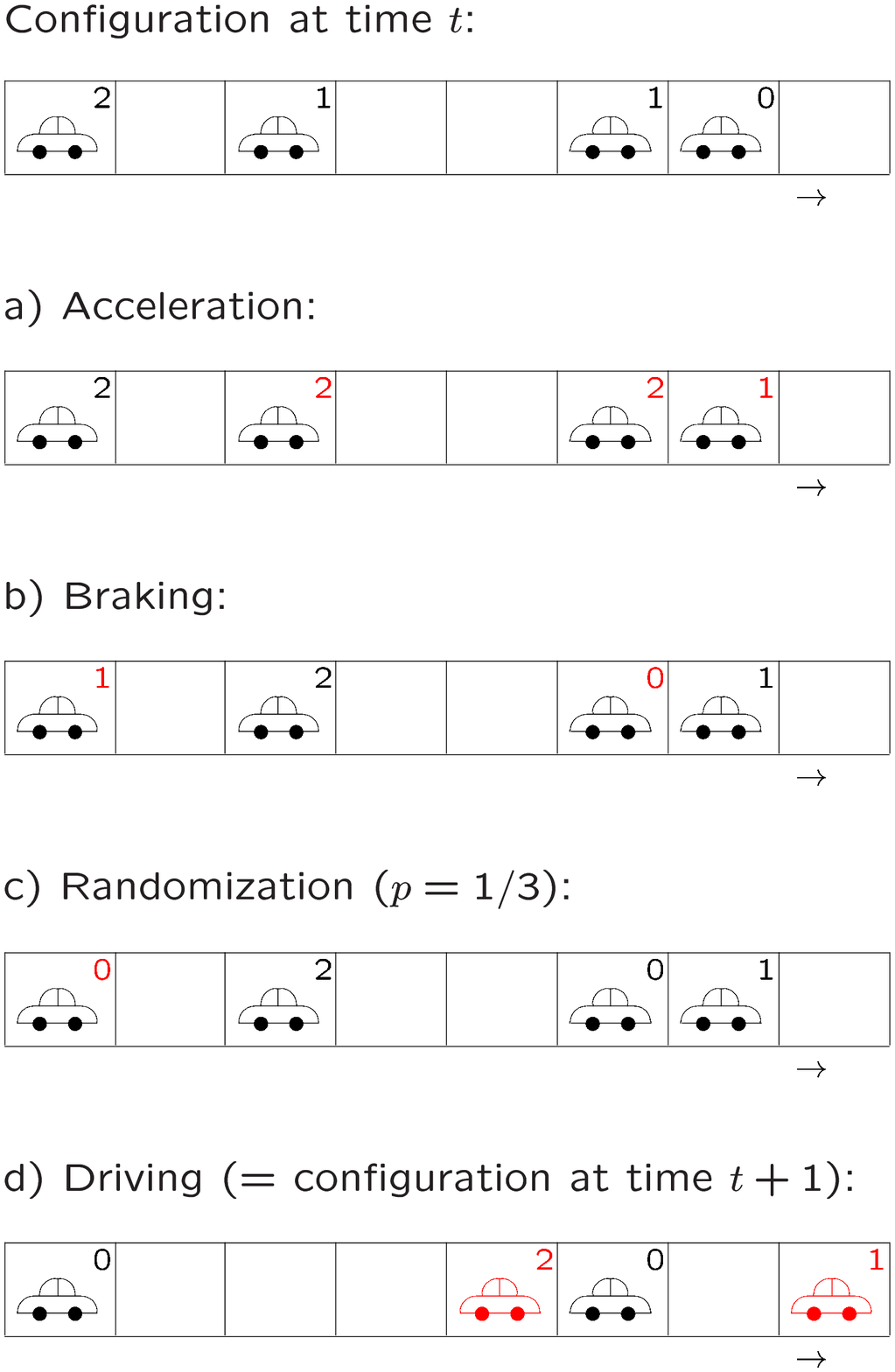,bbllx=40pt,bblly=30pt,bburx=550pt,bbury=760pt,height=12cm}}
\caption{\protect{Step-by-step example for the application of the
update rules. We have assumed $V_{max} = 2$ and $p = 1/3$. Therefore
on average one third of the cars qualifying will slow down in the
randomization step.}}
\label{fig2}
\end{figure}

The update scheme of the NaSch model is illustrated with a simple 
example in fig.\ref{fig2}.

\begin{figure}[h]
 \centerline{\psfig{figure=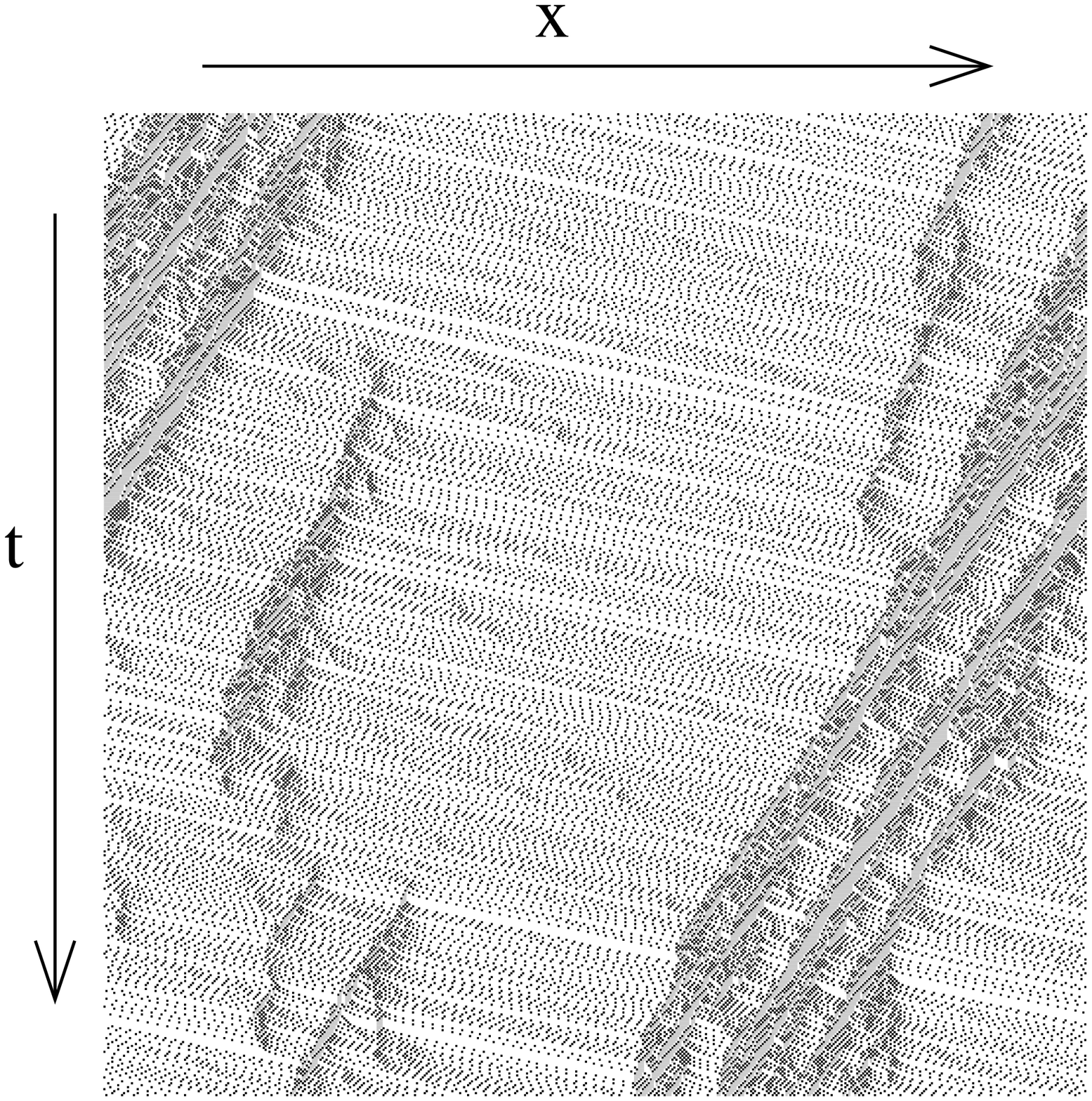,bbllx=10pt,bblly=80pt,bburx=575pt,bbury=680pt,height=4.5cm}
\psfig{figure=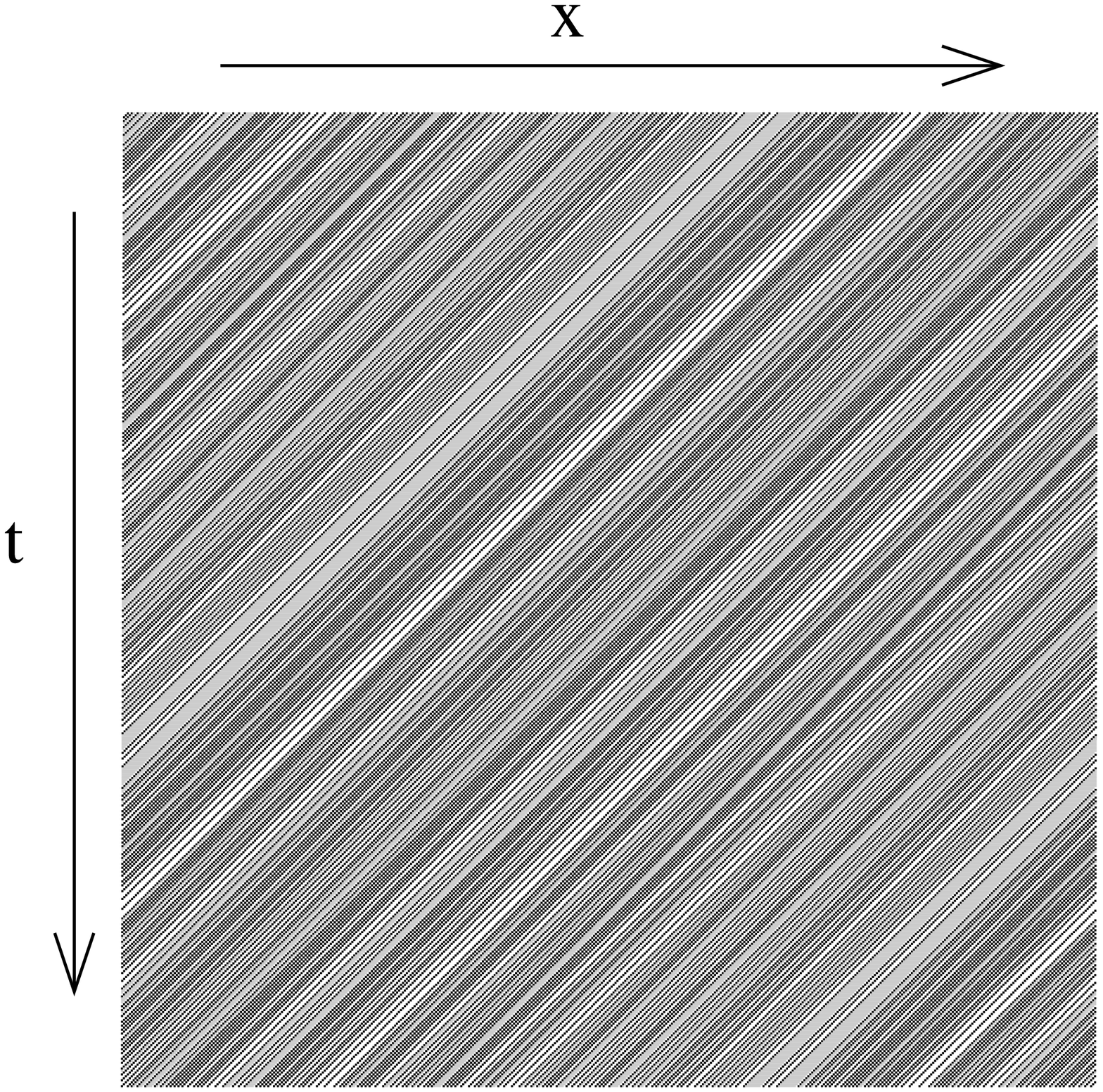,bbllx=10pt,bblly=80pt,bburx=575pt,bbury=680pt,height=4.5cm}}
\caption{\protect{Typical space-time diagrams of the NaSch model 
with $V_{max} = 5$ and (a) $p = 0.25, c = 0.20$, (b) $p = 0.0, c = 0.5$. 
Each horizontal row of dots represents the instantaneous positions of 
the vehicles moving towards right while the successive rows of dots 
represent the positions of the same vehicles at the successive time steps. }}
\label{fig3}
\end{figure}

Space-time diagrams showing the time evolutions of the NaSch model 
demonstrate that no jam is present at sufficiently low densities, 
but spontaneous fluctuations give rise to traffic jams at higher 
densities (fig.3(a)). From the fig.3(b) it should be obvious that 
the {\it intrinsic stochasticity} of the dynamics\cite{ns}, 
arising from non-zero $p$, is essential for triggering the jams 
~\cite{ns,nh}. 

The use of {\it parallel} dynamics is also important. In contrast 
to a random sequential update, it can lead to a chain of 
overreactions. Suppose, a vehicle slows down due the randomization 
step. If the density of vehicles is large enough this might force 
the following vehicle also to brake in the deceleration step. In 
addition, if $p$ is not too small, it might brake even further in 
Step 3. Eventually this can lead to the stopping of a vehicle, 
thus creating a jam. This mechanism of spontaneous jam formation 
is rather realistic and cannot be modelled by the random sequential 
update.

\subsection{Relation between the NaSch model and ASEP}

In the NaSch model with $V_{max} = 1$ every vehicle moves forward 
with probability $q = 1-p$ in the time step $t+1$ if the site 
immediately in front of it were empty at the time step $t$; this, 
is similar to the fully asymmetric simple exclusion process 
({\bf ASEP})~\cite{spohn,sz,gs} where a randomly chosen particle 
can move forward with probablity $q$ if the site immediately in 
front is empty. But, updating is done in parallel in the NaSch model 
whereas that in the ASEP is done in a random sequential manner. 
Nevertheless, the special case of $V_{max} = 1$ for the NaSch model 
achieves special importance from the fact that so far it has been 
possible to derive exact analytical results for the NaSch model only 
in the special limits  (a) $V_{max} = 1$ and arbitrary $p$ and 
(b) $p = 0$ and arbitrary $V_{max}$. 

\subsection{NaSch model in the deterministic limits}

If $p = 0$, the system can self-organize so that at low densities 
every vehicle can move with $V_{max}$ and the corresponding flux 
is $cV_{max}$; this is, however, possible only if enough empty 
cells are available in front of every vehicle, i.e., for 
$ c \leq c_m^{det} = (V_{max} + 1)^{-1}$ and the corresponding 
maximum flux is $J_{max}^{det} = V_{max}/(V_{max} + 1)$. On 
the other hand, for $c > c_m^{det}$, the flow is limited by 
the density of holes. Hence, the  fundamental diagram in the 
deterministic limit $p = 0$ of the NaSch model (for any arbitrary 
$V_{max}$) is given by the {\it exact} expression  
$J = min(c V_{max}, (1 - c))$.
 
Aren't the properties of the NaSch model with maximum allowed 
speed $V_{max}$, in the deterministic limit $p = 1$, exactly 
identical to those of the same model with maximum allowed speed 
$V_{max} - 1$? The answer to the question posed above is: NO; 
if $p = 1$, all random initial states lead to $J = 0$ in the 
stationay state of the NaSch model irrespective of $V_{max}$ and $c$! 

\subsection{Analytical Theory for the NaSch Model} 

In the "site-oriented" theories one describes the state of the 
finite system of length $L$ by completely specifying the state of 
each {\it site}. In contrast, in the "car-oriented" theories the 
state of the traffic system is described by specifying the positions 
and speeds of all the $N$ vehicles in the system. In the naive 
mean-field approximation one treats the probabilities of occupation 
of the lattice sites as independent of each other. In this 
approximation, for example, the steady-state flux for the NaSch 
model with $V_{max} = 1$ and periodic boundary conditions, one 
gets~\cite{ssni} 
\begin{equation}
J = q c (1-c) 
\label{eq-mfflux}
\end{equation}
It turns out\cite{ssni} that the naive mean-field theory 
underestimates the flux for all $V_{max}$. Curiously, if instead of 
parallel updating one uses the random sequential updating, the 
NaSch model with $V_{max} = 1$ reduces to the ASEP for which the 
equation (\ref{eq-mfflux}) is known to be the {\it exact} 
expression for the corresponding flux (see, e.g., \cite{ns})!

What are the reasons for these differences arising from parallel 
updating and random sequential updating? There are "garden of Eden" 
(GoE) states (dynamically forbidden states) ~\cite{ss98} of the NaSch 
model which cannot be reached by the parallel updating whereas no 
state is dynamically forbidden if the updating is done in a random 
sequential manner. 

\begin{figure}[h]
\centerline{\psfig{figure=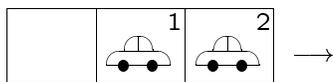,bbllx=50pt,bblly=420pt,bburx=550pt,bbury=540pt,height=2cm}}
\caption{A GoE state for the NaSch model with $V_{max} \geq 2$.}
\label{fig4}
\end{figure}

For example, the configuration shown in fig.4 is a GoE state{\footnote{ 
The configuration shown in fig.1 is also a GoE state!}} because it could 
occur at time $t$ only if the two vehicles occupied the same cell 
simultaneously at time $t - 1$. The naive mean-field theory mentioned above 
does not exclude the GoE states. The exact expression, given in the 
next subsection, for the flux in the steady-state of the NaSch model 
with $V_{max} = 1$ can be derived by merely excluding these states 
from consideration in the naive mean-field theory~\cite{ss98}, thereby 
indicating that the only source of correlation in this case is the 
parallel updating. But, for $V_{max} > 1$, there are other sources of 
correlation because of which exclusion of the GoE states merely 
improves the naive mean-field estimate of the flux but does not yield exact 
results\cite{ss98}. 

A systematic improvement of the naive mean-field theory of the NaSch model 
has been achieved by incorporating short-ranged correlations through 
cluster approximations. We define a 
$n$-cluster to be a collection of $n$ successive sites. 
In the general $n$-cluster approximation, one divides the lattice 
into "clusters" of length $n$ such that two neighbouring clusters 
have $n-1$ sites in common (see fig.5). 
\begin{figure}[h]
 \centerline{\psfig{figure=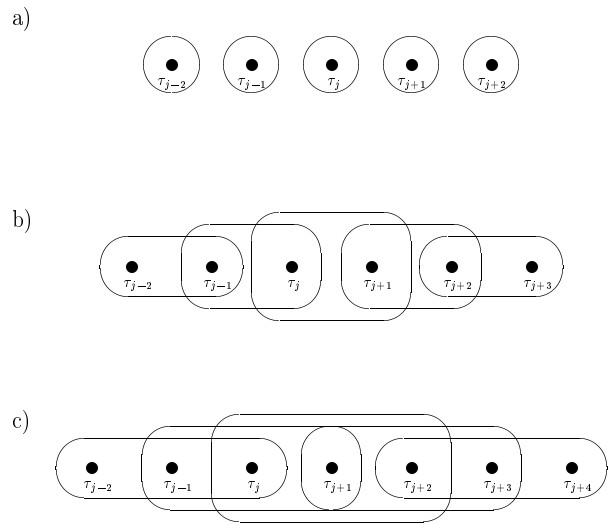,bbllx=80pt,bblly=285pt,bburx=560pt,bbury=740pt,height=8.5cm}}
\caption{\protect{Decomposition of a lattice into (a) 1-clusters
(b) 2-clusters and (c) 3-clusters in the cluster-theoretic approach 
to the NaSch model. }}
\label{fig5}
\end{figure}

If $n = 1$, then the $1$-cluster approximation can be regarded as 
the naive mean-field approximation. You can easily verify, for example, 
in the special case of $V_{max} = 1$, that the state of the 2-cluster 
at time $t+1$ depends on the state of the 4-cluster at time $t$, 
which, in turn, depends on the state of a larger cluster at time $t-1$ 
and, so on. Therefore, one needs to make an approximation to 
truncate this hierarchy in a sensible manner. For example, in the 
2-cluster approximation for the NaSch model with $V_{max} = 1$, the 
4-cluster probabilities are approximated in terms of an appropriate 
product of 2-cluster probabilities. Thus, in the $n$-cluster 
approximation\cite{ssni} a cluster of $n$ neighbouring cells are 
treated exactly and the cluster is coupled to the rest of the system 
in a self-consistent way. 

Carrying out the 2-cluster calculation\cite{ssni} for $V_{max} = 1$ 
one not only finds an effective particle-hole attraction 
(particle-particle repulsion), but also obtains the exact result 
\begin{equation}
J(c,p) = \frac{1}{2}[1 - \sqrt{1- 4qc(1-c)}] 
\label{fl-2cl}
\end{equation} 
for the corresponding flux. But one gets only approximate results 
from the 2-cluster calculations for all $V_{max} > 1$ (see \cite{shad99} 
for higher order cluster calculations for $V_{max} = 2$ and comparison 
with computer simulation data).
\begin{figure}[h]
 \centerline{\psfig{figure=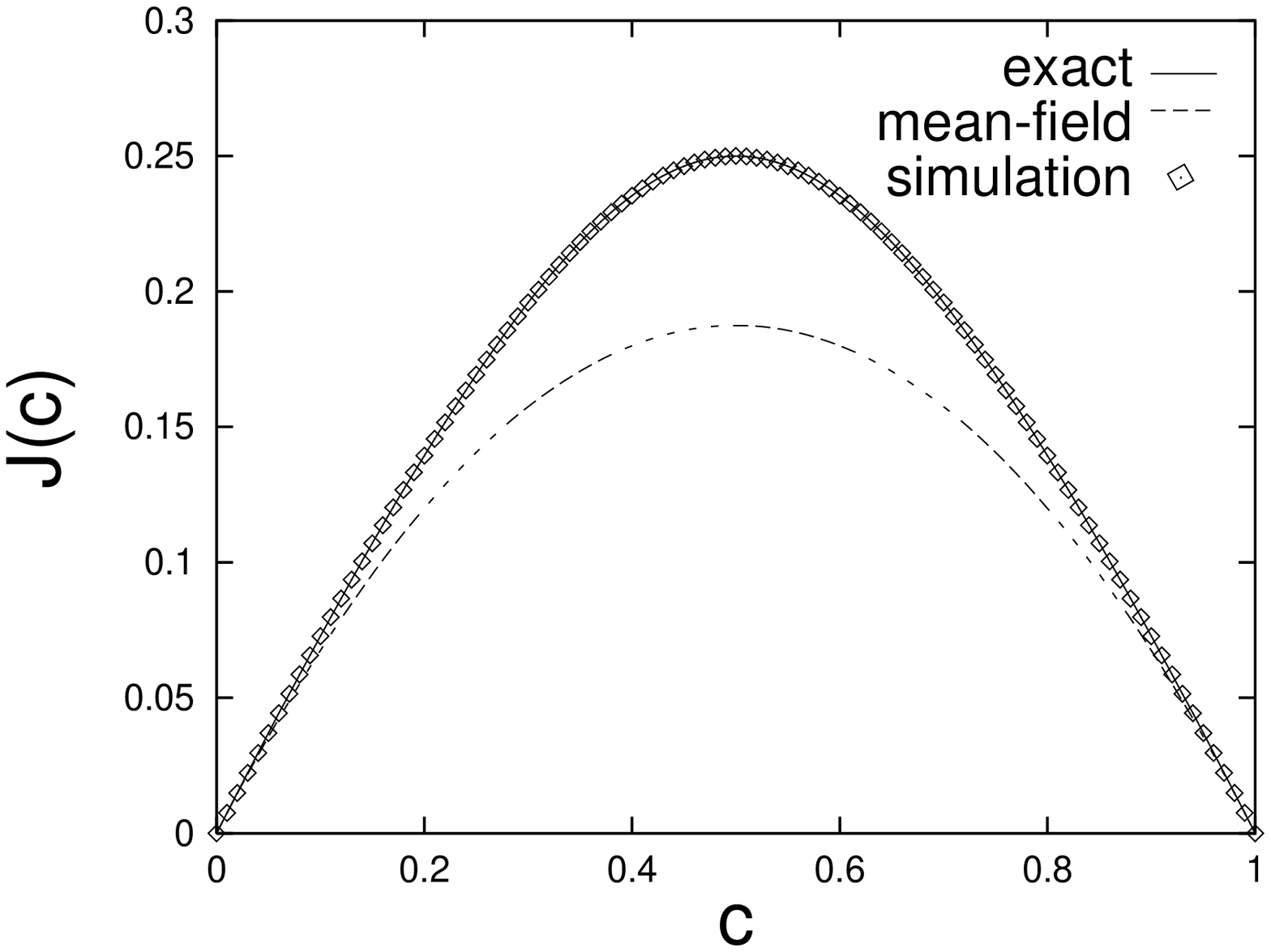,bbllx=50pt,bblly=50pt,bburx=550pt,bbury=400pt,height=3cm}
\psfig{figure=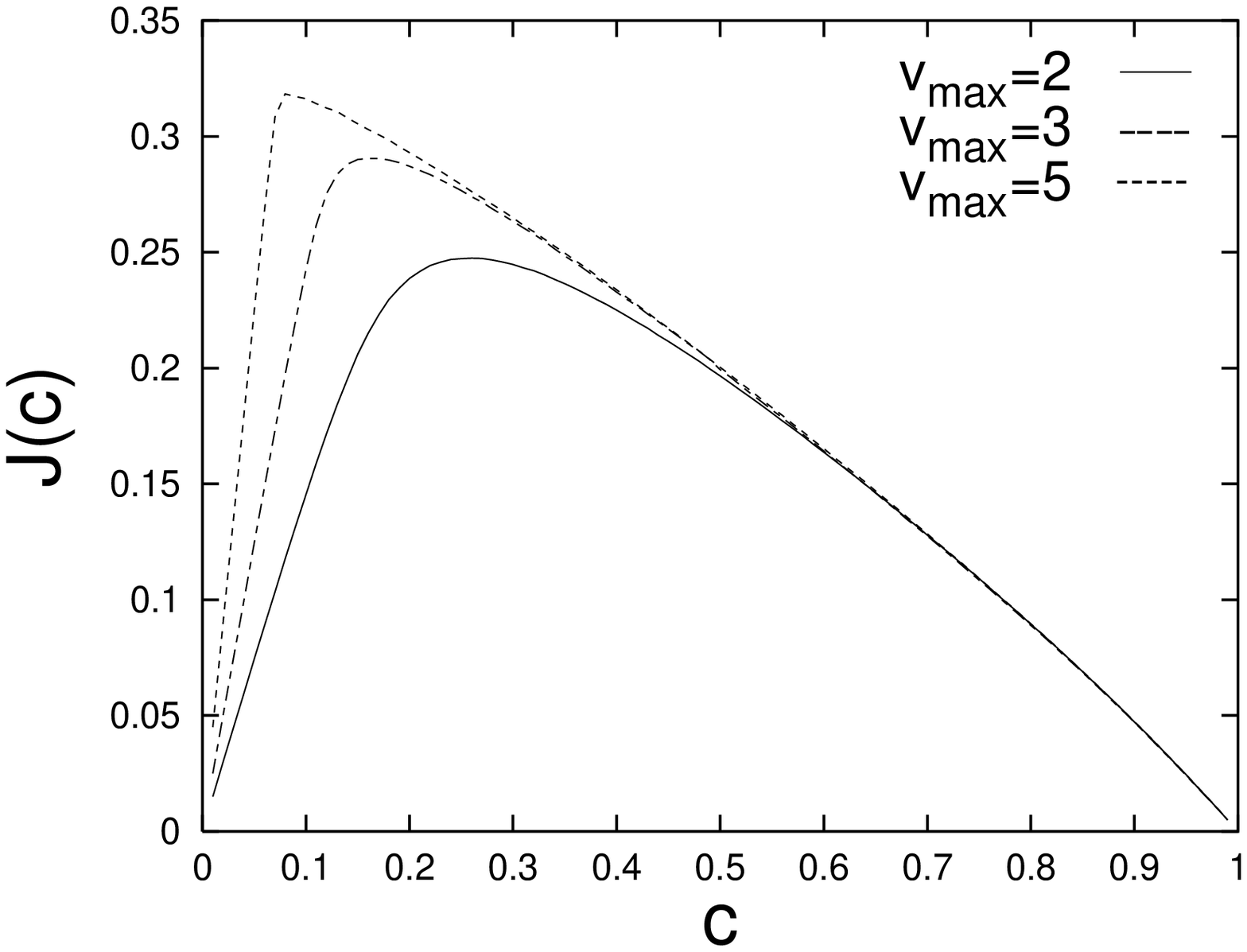,bbllx=50pt,bblly=50pt,bburx=550pt,bbury=400pt,height=3cm}}
\caption{\protect{The fundamental diagram in the NaSch model 
for (a) $V_{max} = 1$ and (b) $V_{max} > 1$, both for $p = 0.25$. 
The data for all $V_{max} > 1$ have been obtained through computer 
simulations. }}
\label{fig6}
\end{figure}

Let us explain the physical origin of the generic shape of the 
fundamental diagrams shown in fig.6.  At sufficiently low density 
of vehicles, practically "free flow" takes place whereas at higher 
densities traffic becomes "congested" and traffic jams occur. So 
long as $c$ is sufficiently small, the average speed 
$\langle V \rangle$ is practically independent of $c$ as the vehicles 
are too far apart to interact mutually. However, a faster monotonic 
decrease of $\langle V \rangle$ with increasing $c$ takes place when 
the forward movement of the vehicles is strongly hindred by others 
because of the reduction in the average separation between them. 
Because of this trend of variation of $\langle V \rangle$ with $c$, 
the flux $J = \langle c V \rangle$ exhibits a maximum \cite{may90} 
at $c_m$; for $c < c_m$, increasing $c$ leads to increasing $J$ 
whereas for $c > c_m$ sharp decrease of $\langle V \rangle$ with 
increase of $c$ leads to the overall decrease of $J$.  

An interesting feature of the expression (\ref{fl-2cl}) is that 
the flux is invariant under charge conjugation, i.e., under the 
operation $c \rightarrow (1-c)$ which interchanges particles and 
holes. Therefore, the fundamental diagram is symmetric about 
$c = 1/2$ when $V_{max} = 1$ (see fig.6a). Although this symmetry 
breaks down for all $V_{max} > 1$ (see fig.6b), the corresponding 
fundamental diagrams appear more realistic. Moreover, for given 
$p$, the magnitude of $c_m$ decreases with increasing $V_{max}$ 
as the higher is the $V_{max}$ the longer is the effective range 
of interaction of the vehicles (see fig.6b). Furthermore, for 
$V_{max} = 1$, flux merely decreases with increasing $p$  
(see eqn(\ref{fl-2cl})), but remains symmetric about $c = 1/2 =c_m$. 
On the other hand, for all $V_{max} > 1$, increasing $p$ not only 
leads to smaller flux but also lowers $c_m$.

\subsection{Spatio-temporal organization of vehicles} 

The distance from a selected point on the lead vehicle to the same 
point on the following vehicle is defined as the {\it distance-headway} 
({\bf DH}) \cite{may90}. In order to get information on the spatial 
organization of the vehicles, one can calculate the DH distribution 
${\cal P}_{dh}(\Delta X)$ by following either a site-oriented approach 
\cite{chow97a} or a car-oriented approach\cite{ss97} if 
$\Delta X_j = X_j - X_{j-1}$, i.e., if the number of empty lattice 
sites in front of the $j$-th vehicle is identified as the corresponding 
DH. At moderately high densities, ${\cal P}_{dh}(\Delta X)$ exhibits 
two peaks; the  peak at $\Delta X = 1$ is caused by the jammed vehicles 
while that at a larger $\Delta X$ corresponds to the most probable DH 
in the free-flowing regions. 

The {\it time-headway} is defined as the time interval between the 
departures (or arrivals) of two successive vehicles recorded by a 
detector placed at a fixed position on the highway \cite{may90}. The 
time-headway distribution contains information on the 
temporal organization.  Suppose, ${\cal P}_m(t_1)$ is the probability 
that the following vehicle takes time $t_1$ to reach the detector, 
moving from its initial position where it was located when the 
leading vehicle just left the detector site. Suppose, after reaching 
the detector site, the following vehicle waits there for $\tau - t_1$ 
time steps, either because of the presence of another vehicle in front 
of it or because of its own random braking; the probability for this 
event is denoted by $Q(\tau-t_1|t_1)$. The distribution 
${\cal P}_{th}(\tau)$, of the time-headway $\tau$, can be obtained 
from \cite{chow98a,ghosh} 
${\cal P}_{th}(\tau) = \sum_{t_1=1}^{\tau-1} {\cal P}_m(t_1) Q(\tau-t_1|t_1)$. 
The most-probable time-headway, when plotted against the density, 
exhibits a minimum \cite{ghosh}; this is consistent with the well 
known exact relation $J = 1/T_{av}$ between flux and the average 
time-headway, $T_{av}$.

Is there a phase transition from "free-flowing" to "congested" 
dynamical phase of the NaSch model? No satisfacory order parameter 
has been found so far\cite{eisen,chebani}, except in the 
deterministic limit\cite{vdes}. The possibility of the existence 
of any critical density in the NaSch model is ruled out by the 
observations \cite{eisen,chebani,shad99,kn94} that, for all non-zero 
$p$, (a) the equal-time correlation function decays exponentially 
with separation, and (b) the relaxation time and lifetimes of the 
jams remain finite. This minimal model of highway traffic also 
does not exhibit any first order phase transition and two-phase 
co-existence\cite{chow98a}. 

\subsection{Extensions of the NaSch model and practical applications} 

In recent years some other minimal models of traffic on highways 
have been developed by modifying the updating rules of the NaSch 
model \cite{fukuishi,wang98,helbshrek}. In the cruise control 
limit of the NaSch model \cite{paczus} the randomization step is 
applied only to vehicles which have a velocity $V < V_{max}$ after 
step 2 of the update rule. Vehicles moving with their desired 
velocity $V_{max}$ are not subject to fluctuations. This is exactly 
the effect of a cruise-control which automatically keeps the 
velocity constant at a desired value.  Interestingly, the 
cruise-control limit of the NaSch model exhibits self-organized 
criticality\cite{bak,ddhar}. Besides, a continuum limit of the 
NaSch model has also been considered\cite{krauss}.

The vehicles which come to a stop because of hindrance from the 
leading vehicle may not be able to start as soon as the leading 
vehicle moves out of its way; it may start with a probability 
$q_s < 1$. When such possibilities are incorporated in the NaSch 
model, the "slow-to-start" rules \cite{bjh,tt,barlovic,sss2s,chow99} 
can give rise to metastable states of very high flux and hysteresis 
effects as well as phase separation of the traffic into a 
"free-flowing" phase and a "mega-jam". 

The bottleneck created by quenched disorder of the {\it highway} 
usually slows down traffic and can give rise to jams 
\cite{cv,chow98a} and phase segregation\cite{lebo,barma}. However, 
a different type of quenched disorder, introduced by assigning 
randomly different braking probabilities $p$ to different drivers 
in the NaSch model, can have more dramatic effects\cite{ktitarev,santen99} 
which are reminiscent of "Bose-Einstein-like condensation" 
in the fully ASEP where particle-hopping rates are quenched random 
variables\cite{kf,evans}. In such "Bose-Einstein-like condensed" state 
finite fraction of the empty sites are "condensed" in front of the 
slowest vehicle (i.e., the driver with highest $p$). 

Several attempts have been made to generalize the NaSch model to 
describe traffic on multi-lane highways and to simulate traffic 
on real networks in and around several cities\cite{wolf}. 
For {\it planning and design} of the transportation network 
\cite{nagel99}, for example, in a metropolitan area 
\cite{duis,dallas1,dallas2}, one needs much more than just 
{\it micro-simulation} of how vehicles move on a linear or 
square lattice under a specified set of vehicle-vehicle and 
road-vehicle interactions. For such a simulation, to begin with, 
one needs to specify the roads (including the number of lanes, 
ramps, bottlenecks, etc.) and their intersections. Then, times 
and places of the activities, e.g., working, shopping, etc., 
of individual drivers are planned.  Micro-simulations are 
carried out for all possible different routes to execute 
these plans; the results give informations on the efficiency 
of the different routes and these informations are utilized in 
the designing of the transportation network 
\cite{rickertdipl,nageldiss,nagel99}. Some socio-economic 
questions as well as questions on the environmental impacts 
of the planned transportation infrastructure also need to 
be addressed during such planning and design.

\section{Cellular-automata Models of City-traffic}

\subsection{The Biham-Middleton-Levin model of city traffic and its generalizations} 

In the BML model\cite{bml}, each of the sites of a square lattice 
represent the crossing of a east-west street and a north-south street. 
All the streets parallel to the $\hat{X}$-direction of a Cartesian 
coordinate system are assumed to allow only {\it single-lane} east-bound 
traffic while all those parallel to the $\hat{Y}$-direction allow 
only single-lane north-bound traffic. In the initial state of the 
system, vehicles are randomly distributed among the streets. 
The states of east-bound vehicles are updated in parallel at every 
odd discrete time step whereas those of the north-bound vehicles 
are updated in parallel at every even discrete time step following 
a rule which is a simple extension of the fully ASEP: a vehicle 
moves forward by one lattice spacing if and only if the site in 
front is empty, otherwise the vehicle does not move at that time step. 

Computer simulations demonstrate that a {\it first order} phase 
transition takes place in the BML model at a finite non-vanishing 
density $c_*$, where the average velocity of the vehicles vanishes 
{\it discontinuously} signalling complete jamming; this jamming arises 
from the mutual blocking of the flows of east-bound and north-bound 
traffic at various different crossings\cite{tadaki,gupta}. Note 
that the dynamics of the BML model is fully {\it deterministic} 
and the randomness arises only from the {\it random initial 
conditions} \cite{fuishjp96}. 

As usual, in the naive mean-field approximation one neglects the 
correlations between the occupations of different sites\cite{mmcb95}. 
However, if you are not interested in detailed information on the 
"structure" of the dynamical phases, you can get a mean-field 
estimate of $c_*$ by carrying out a back-of-the-envelope 
calculation\cite{nagajp93,nagapr93,wfh96}. In the symmetric case 
$c_x = c_y$, for which $v_x = v_y = v$, $c = c_* \simeq 0.343$. 

The BML model has been extended to take into account the effects 
of (i) asymmetric distribution of the vehicles\cite{nagajp93}, 
i.e., $c_x \neq c_y$, 
(ii) overpasses or two-level crossings\cite{nagapr93} that are 
represented by specifically identified sites each of which can 
accomodate upto a maximum of two vehicles simultaneously,
(iii) faulty traffic lights\cite{chung95} 
(iv) static hindrances or road blocks or vehicles crashed in traffic 
accident, i.e., stagnant points\cite{acci,gu}, 
(v) stagnant street where the local density $c_s$ of the vehicles 
is initially higher than that in the other streets\cite{stag}
(vi) jam-avoiding drive\cite{jamavoid} of vehicles to a neighbouring 
street, parallel to the original direction, to avoid getting blocked 
by other vehicles in front,
(vii) turning of the vehicles from east-bound (north-bound) to 
north-bound (east-bound) streets\cite{turn}. 
(viii) a single north-bound street cutting across east-bound streets
\cite{nagaseno}
(ix) more realistic description of junctions of perpendicular 
streets \cite{freund,chopard},  
(x) green-waves \cite{greenwave}.

\subsection{Marriage of NaSch and BML models} 

At first sight the BML model may appear very unrealistic because the 
vehicles seem to hop from one crossing to the next. However, it 
may not appear so unrealistic if each unit of discrete 
time interval in the BML model is interpreted as the time for 
which the traffic lights remain green (or red) before switching 
red (or green) simultaneously in a synchronized manner, and over 
that time scale each vehicle, which faces a green signal, gets an 
opportunity to move from $j$-th crossing to the $j+1$-th (or, more 
generally\cite{fukui}, to the $j+r$-th where $r > 1$). 

However, if one wants to develope a more detailed "fine-grained" 
description then one must first decorate each bond\cite{horiguchi} 
with $D-1$ ($D > 1$) sites to represent $D-1$ cells in between each 
pair of successive crossings thereby modelling each segment of 
the streets in between successive crossings in the same manner in   
which the entire highway is modelled in the NaSch model. Then, one 
can follow the prescriptions of the NaSch model for describing the 
positions, speeds and accelerations of the vehicles \cite{chopard,simon} 
as well as for taking into account the interactions among the 
vehicles moving along the same street. Moreover, one should flip 
the color of the signal periodically at regular interval of $T$ 
($T >> 1$) time steps where, during each unit of the discrete time 
interval every vehicle facing green signal should get an opportunity 
to move forward from one cell to the next. Such a CA model of traffic 
in cities has, indeed, been proposed very recently ~\cite{cs} 
where  the rules of updating have been formulated in such a way 
that, (a) a vehicle approaching a crossing can keep moving, even 
when the signal is red, until it reaches a site immediately in 
front of which there is either a halting vehicle or a crossing; 
and (b) no grid-locking would occur in the absence of random 
braking. 

\begin{figure}[hbt]
\epsfxsize=\columnwidth\epsfbox{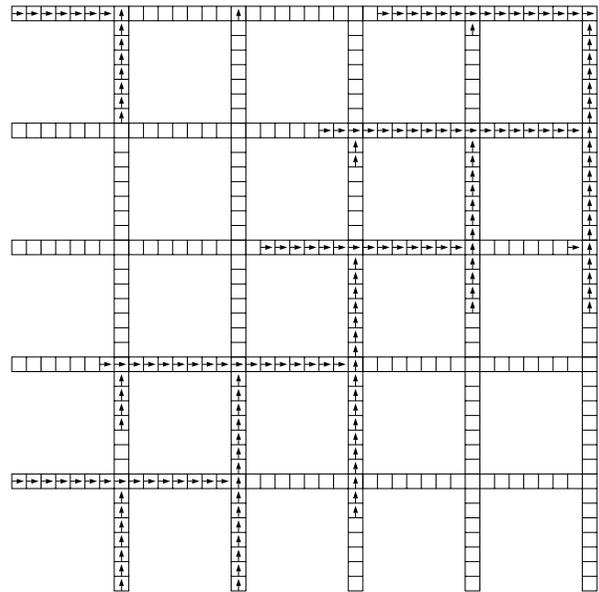}
\caption{A typical jammed configuration of the vehicles.
The east-bound and north-bound vehicles are represented by the
symbols $\rightarrow$ and $\uparrow$, respectively. ($N = 5, D = 8$)}
\label{fig7}
\end{figure}

A phase transition from the "free-flowing" dynamical phase to the 
completely "jammed" phase has been observed in this model at a vehicle 
density which depends on $D$ and $T$. The intrinsic stochasticity 
of the dynamics, which triggers the onset of jamming, is similar to 
that in the NaSch model, while the phenomenon of complete jamming through 
self-organization as well as the final jammed configurations (fig.7) 
are similar to those in the BML model. This model also provides a 
reasonable time-dependence of the average speeds of the vehicles in 
the "free-flowing" phase\cite{cs}. 

\section{Relation with other systems and phenomena} 

You must have noticed in the earlier sections that some of the 
models of traffic are non-trivial generalizations or extensions of 
the ASEP, the simplest of the {\it driven-dissipative} systems 
which are of current interest in non-equilibrium statistical 
mechanics\cite{sz}. Some similarities between these systems and 
a dynamical model of protein synthesis has been pointed out\cite{shubio}.
Another driven-dissipative system, which is also receiving wide 
attention of physicists in recent years, is the granular material 
flowing through a pipe\cite{proc1,proc2}. There are some 
superficial similarities between the clustering of vehicles on 
a highway and particle-particle (and particle-cluster) aggregation 
process\cite{ben-naim}. 

The NaSch model with $V_{max} = 1$ can be mapped onto stochastic 
growth models of one-dimensional surfaces in a two-dimensional 
medium. Particle (hole) movement to the right (left) correspond to 
local forward growth of the surface via particle deposition. 
In this scenario a particle evaporation would correspond to 
a particle (hole) movement to the left (right) which is not 
allowed in the NaSch model. It is worth pointing out that 
any quenched disorder in the rate of hopping between two 
adjacent sites would correspond to {\it columnar} quenched 
disorder in the growth rate for the surface\cite{barma}. 

Inspired by the recent success in theoretical studies of traffic, 
some studies of information traffic on the computer network 
(internet) have also been carried out\cite{csabai,ohira,taka}. 

\section{Summary and conclusion:} 

Nowadays the tools of statistical mechanics are increasingly 
being used to study self-organization and emergent collective 
behaviour of {\it complex systems} many of which, including 
vehicular traffic, fall outside the traditional domain of 
physical systems. However, as we have shown in this article, 
a strong theoretical foundation of traffic science can be 
built on the basic principles of statistical mechanics. In 
this brief review we have focussed attention mainly on the 
progress made in the recent years using "particle-hopping" 
models, formulated in terms of cellular automata, and compared 
these with several other similar systems.

\noindent{\bf Acknowledgements:} It is our pleasure to thank 
R. Barlovic, J.G. Brankov, B. Eisenbl\"atter, K. Ghosh, N. Ito,
K. Klauck, W. Knospe, D. Ktitarev, A. Majumdar, K. Nagel, V.B.
Priezzhev, M. Schreckenberg, A. Pasupathy, S. Sinha, R.B.
Stinchcombe and D.E. Wolf for enjoyable collaborations the results
of some of which have been reviewed here. We also thank M. Barma,
J. Kertesz, J. Krug, G. Sch\"utz, D. Stauffer and J. Zittartz for
useful discussions and encouragements.
This work is supported by SFB341 K\"oln-Aachen-J\"ulich.



\begin{references}

\item[$^{\ast}$] This is a modified and shortened version of a longer detailed
review article to be published elsewhere.

\item[$^{\dag}$]Permanent address.


\bibitem{hg} R. Herman and K. Gardels, Sci.Am. {\bf 209}(6), 35 (1963)

\bibitem{gazis} D.C. Gazis, Science, {\bf 157}, 273 (1967)

\bibitem{ph} I. Prigogine and R. Herman, {\sl Kinetic Theory of Vehicular Traffi
c} (Elsevier, Amsterdam, 1971)

\bibitem{proc1} D.E. Wolf, M. Schreckenberg and A. Bachem (eds.) {\sl Traffic and Granular Flow} (World Scientific, Singapore, 1996) 

\bibitem{proc2} D.E. Wolf and M. Schreckenberg (eds.) {\sl Traffic and Granular Flow} (Springer, Singapore, 1998) 

\bibitem{helbsci} D. Helbing and M. Treiber, Science {\bf 282}, 2002 (1998)

\bibitem{wolfram} S. Wolfram, {\it Theory and Applications of Cellular Automata}, (World Scientific, 1986).


\bibitem{ns} K. Nagel and M. Schreckenberg, J. Physique I, {\bf2},
2221 (1992).

\bibitem{bml} O. Biham, A.A. Middleton and D. Levine, Phys. Rev. A {\bf 46},
R6124 (1992)

\bibitem{may90} A.D. May, {\sl Traffic Flow Fundamentals} (Prentice-Hall, 1990)

\bibitem{kerner97} B.S. Kerner, Phys. Rev. Lett {\bf 81}, 3797 (1998) 

\bibitem{lw} M.J. Lighthill and G.B. Whitham, Proc. Roy. Soc. Lond. A
{\bf 229}, 281 (1955)

\bibitem{kernetal} B.S. Kerner, S.L. Klenov and P. Konh\"auser, Phys. Rev. E {\bf 56}, 4200 (1997)

\bibitem{hongetal} D.C. Hong and S. Yue, Phys. Rev. E {\bf 58}, 4763 (1998).

\bibitem{leeetal} H.Y. Lee, H.-W. Lee and D. Kim, Phys. Rev. Lett. {\bf 81}, 1130 (1998)

\bibitem{helbkin} D. Helbing and M. Treiber, Phys. Rev. Lett.{\bf 81}, 3042 (1998) 

\bibitem{wagkin} C. Wagner, C. Hoffmann, R. Sollacher, J. Wagenhuber and B. Sch\"urmann, Phys. Rev. E {\bf 54}, 5073 (1996)

\bibitem{nagakin} T. Nagatani, Physica A {\bf 237}, 67 (1997) 
 
\bibitem{bandoetal} K. Nakanishi, K. Itoh, Y. Igarashi and M. Bando, Phys. Rev. E {\bf 55}, 6519 (1997)

\bibitem{komatsu} T.S. Komatsu and S. Sasa, Phys. Rev. E {\bf 52}, 5574 (1995) 

\bibitem{hayanaka} H. Hayakawa and K. Nakanishi, Phys. Rev. E {\bf 57}, 3839 (1998) 

\bibitem{mason} A.D. Mason and A.W. Woods, Phys. Rev. E {\bf 55}, 2203 (1997) 

\bibitem{sugiyama} Y. Sugiyama and H. Yamada, Phys. Rev. E {\bf 55}, 7749 (1997)  

\bibitem{nagatani} T. Nagatani, and K. Nakanishi, Phys. Rev. E {\bf 57}, 6415 (1998)  

\bibitem{wagner} C. Wagner, Physica A {\bf 260}, 218 (1998)

\bibitem{nh} K. Nagel and H.J. Herrmann, Physica A {\bf 199}, 254 (1993)

\bibitem{spohn} H. Spohn, {\it Large scale dynamics of interacting particles} (Springer, 1991) 

\bibitem{sz} B. Schmittmann and R.K.P. Zia, in:{\it Phase Transitions and
Critical Phenomena}, eds. C. Domb and J.L. Lebowitz vol.17 (Academic Press,
1995)

\bibitem{gs} G. Sch\"utz, in: {\it Phase transitions and critical phenomena}, eds. C. Domb and J.L. Lebowitz (to appear) 

\bibitem{ssni} M. Schreckenberg, A. Schadschneider, K. Nagel and N. Ito, Phys. Rev. E {\bf 51}, 2939 (1995)

\bibitem{ss98} A.\ Schadschneider and M.\ Schreckenberg,
  J.\ Phys.\ A {\bf 31}, L225 (1998)

\bibitem{shad99} A. Schadschneider, Eur. Phys. J. B (1999) (in press)

\bibitem{ss97} A.\ Schadschneider and M.\ Schreckenberg,
  J.\ Phys.\ A {\bf 30}, L69 (1997).

\bibitem{chow97a} D. Chowdhury, A. Majumdar, K. Ghosh, S. Sinha and R.B. Stinchcombe , Physica A {\bf 246}, 471 (1997).

\bibitem{chow98a} D. Chowdhury, A. Pasupathy and S. Sinha, Eur. Phys. J. B {\bf 5}, 781 (1998) 

\bibitem{ghosh}  K. Ghosh, A. Majumdar and D. Chowdhury, Phys. Rev. E {\bf 58}, 4012 (1998)

\bibitem{eisen} B. Eisenbl\"atter, L. Santen, A. Schadschneider and M. Schreckenberg, Phys. Rev. E {\bf 57}, 1309 (1998).

\bibitem{chebani} S. Cheybani, J. Kertesz and M. Schreckenberg, J. Phys. A {\bf 31}, 9787 (1998).

\bibitem{vdes} L.C.Q. Vilar and A.M.C. De souza, Physica A {\bf 211}, 84 (1994)

\bibitem{kn94} K. Nagel, Int. J. Mod. Phys. C, {\bf 5}, 567 (1994) 

\bibitem{fukuishi} M. Fukui and Y. Ishibashi, J. Phys. Soc. Jap. {\bf 65}, 1868 (1996)

\bibitem{wang98} B.H. Wang, Y.R. Kwong and P.M. Hui, Physica A {\bf 254}, 122 (1998)

\bibitem{helbshrek} D. Helbing and M. Schreckenberg, Phys. Rev. E {\bf 59}, R2505 (1999)

\bibitem{paczus} K. Nagel and M. Paczuski, Phys. Rev. E {\bf 51}, 2909 (1995)

\bibitem{bak} P. Bak, {\it Self-Organized Criticality: Why Nature is Complex} (Springer, 1996) 

\bibitem{ddhar} D. Dhar, Physica A {263}, 4 (1999) and references therein

\bibitem{krauss} S. Krauss, P. Wagner and C. Gawron, Phys. Rev. E {\bf 54}, 3707 (1996); {\bf 55}, 5597 (1997) 

\bibitem{bjh}S.C. Benjamin, N.F. Johnson and P.M. Hui, J. Phys. A {\bf 29}, 311
9 (1996)

\bibitem{tt}M. Takayasu and H. Takayasu, Fractals, {\bf 1}, 860 (1993)

\bibitem{barlovic} R. Barlovic, L. Santen, A. Schadschneider and M. Schreckenberg, Eur. Phys. J. {\bf 5}, 793 (1998) 

\bibitem{sss2s} A. Schadschneider and M. Schreckenberg, Ann. der Phys. {\bf 6}, 541 (1997).

\bibitem{chow99} D. Chowdhury, L. Santen, A. Schadschneider, S. Sinha and A. Pasupathy, J. Phys. A {\bf 32}, 3229 (1999). 

\bibitem{cv} Z. Csahok and T. Vicsek, J. Phys. A {\bf 27}, L591 (1994)

\bibitem{lebo} S.A. Janowsky and J.L. Lebowitz, J. Stat. Phys. {\bf 77}, 35 (1994) 

\bibitem{barma} G. Tripathy and M. Barma, Phys. Rev. Lett. {\bf 78}, 3039 (1997)

\bibitem{ktitarev} D. Ktitarev, D. Chowdhury and D.E. Wolf, J. Phys. A {\bf 30}, L221 (1997)

\bibitem{santen99} W. Knospe, L. Santen, A. Schadschneider and M. Schreckenberg,  Physica A {\bf 265}, 614 (1999) 

\bibitem{kf} J. Krug and P. A. Ferrari, J. Phys. A {\bf 29}, L465 (1996)

\bibitem{evans} M. R. Evans, Europhys. Lett.{\bf 36}, 13 (1996)   

\bibitem{wolf} D.E. Wolf, Physica A, {\bf 263}, 438 (1999); D. Chowdhury, D. E. Wolf and M. Schreckenberg, Physica A {\bf 235} , 417 (1997).

\bibitem{duis} J.\ Esser, M.\ Schreckenberg: Int.\ J.\ Mod.\ Phys. {\bf C},
in press

\bibitem{dallas1} M.\ Rickert, K.\ Nagel: Int.\ J.\ Mod.\ Phys. {\bf C8},
483 (1997)

\bibitem{dallas2} K.\ Nagel, C.L.\ Barrett: Int.\ J.\ Mod.\ Phys. {\bf C8},
505 (1997)

\bibitem{rickertdipl} N.\ Rickert: Diploma thesis, Cologne University (1994)

\bibitem{nageldiss} K.\ Nagel: Ph.D.\ thesis, Cologne University (1995)

\bibitem{nagel99} K. Nagel, J. Esser and M. Rickert, in: {Annu. Rev. Comp. Phys.} ed. D. Stauffer (1999)

\bibitem{tadaki} S. Tadaki and M. Kikuchi, Phys. Rev. E {\bf 50}, 4564 (1994); J. Phys. Soc. Jap. {\bf 64}, 4504 (1995)

\bibitem{gupta} H.S. Gupta and R.Ramaswamy, J.Phys.A 29, L547 (1996). 

\bibitem{fuishjp96} M. Fukui and Y. Ishibashi, J.Phys.Soc.Jap.65, 1871 
(1996). 

\bibitem{mmcb95} J.M. Molera, F.C. Martinez, J.A. Cuesta and R. Brito, 
Phys. Rev. E, {\bf 51}, 175 (1995) 

\bibitem{nagajp93} T. Nagatani, J.Phys.Soc.Jap.62, 2656(1993) 

\bibitem{nagapr93} T. Nagatani, Phys. Rev. E {\bf 48}, 3290 (1993) 

\bibitem{wfh96} B.H. Wang, Y.F. Woo and P.M. Hui, J.Phys.A29, L31 (1996); J.Phys.Soc.Jap.65, 2345 (1996).  

\bibitem{chung95} K.H. Chung, P.M. Hui and G.Q. Gu, Phys. Rev. E {\bf 51}, 772 (1995)

\bibitem{acci} T. Nagatani, J.Phys.A 26, L1015 (1993); J. Phys. Soc. Jap. {\bf 62}, 1085 (1993) 

\bibitem{gu} G.Q. Gu, K.H. Chung and P.M. Hui, Physica A 217, 339 (1995). 

\bibitem{stag} T. Nagatani, Physica A 198, 108 (1993) 

\bibitem{jamavoid} T. Nagatani, J. Phys. Soc. Jap. {\bf 64}, 1421 (1995) 

\bibitem{turn} F.C. Martines, J.A. Cuesta, J.M. Molera and R. Brito, Phys. Rev. E{\bf 51}, R835 (1995); T. Nagatani, J. Phys. Soc. Jap. {\bf 63}, 1228 (1994) 

\bibitem{nagaseno} T. Nagatani and T. Seno, Physica A {\bf 207}, 574 (1994) 

\bibitem{freund} J. Freund and T. P\"oschel, Physica A {\bf 219}, 95 (1995)

\bibitem{chopard} B. Chopard, P.O. Luthi and P.A. Queloz, J. Phys. A {\bf 29}, 2325 (1996)

\bibitem{greenwave} J. T\"or\"ok and J. Kertesz, Physica A {\bf 231}, 515 (1996)

\bibitem{fukui} M. Fukui, H. Oikawa and Y. Ishibashi, J. Phys.Soc.Jpn.{\bf 65}, 2514 (1996) 

\bibitem{horiguchi} T. Horiguchi and T. Sakakibara, Physica A {\bf 252}, 388 (1998)

\bibitem{simon} P.M. Simon and K. Nagel, Phys. Rev. E {\bf 58}, 1286 (1998)

\bibitem{cs} D. Chowdhury and A. Schadschneider, Phys. Rev. E {\bf 59}, R1311 (1999); D. Chowdhury, K. Klauck, L. Santen, A. Schadschneider and J. Zittartz (to be published)

\bibitem{shubio} G.M. Sch\"utz, Int. J. Mod.Phys.B{\bf 11}, 197 (1997) 

\bibitem{ben-naim} E. Ben-Naim and P.L. Krapivsky, Phys. Rev. E {\bf 56}, 6680 (1997)

\bibitem{csabai} I. Csabai, J. Phys. A. {\bf 27}, L417 (1994). 

\bibitem{ohira} T. Ohira and R. Sawatari, Phys. Rev. E {\bf 58}, 193 (1998)

\bibitem{taka} M. Takayasu, A. Yu Tretyakov, K. Fukuda and H. Takayasu, in ref.\cite{proc2}.

\end{references}
\end{document}